# A Universal Approach to Electrostatically-Blind Quantitative Piezoresponse Force Microscopy

Jason P. Killgore, Larry Robins, Liam Collins

Applied Chemicals and Materials Division, National Institute of Standards and Technology, Boulder, CO, USA

Center for Nanophase Materials Sciences, Oak Ridge National Laboratory, Oak Ridge, TN, USA

Contact: Jason.Killgore@NIST.gov

**Abstract**

The presence of electrostatic forces and associated artifacts complicates the interpretation of piezoresponse force microscopy (PFM) and electrochemical strain microscopy (ESM). Eliminating these artifacts provides an opportunity for precisely mapping domain wall structures and dynamics, accurately quantifying local piezoelectric coupling coefficients, and reliably investigating hysteretic processes at the single nanometer scale to determine properties and mechanisms which underly important applications including computing, batteries and biology. Here we exploit the existence of an electrostatic blind spot (ESBS) along the length of the cantilever, due to the distributed nature of the electrostatic force, which can be universally used to separate unwanted long range electrostatic contributions from short range electromechanical responses of interest. The results of ESBS-PFM are compared to state-of-the-art interferometric displacement sensing PFM, showing excellent agreement above their respective noise floors. Ultimately, ESBS-PFM allows for absolute quantification of piezoelectric coupling coefficients independent of probe, lab or experimental conditions. As such, we expect the widespread adoption of EBSB-PFM to be a paradigm shift in the quantification of nanoscale electromechanics.

**Introduction**

Since its invention in 1986[1], the atomic force microscope (AFM) has offered unparalleled opportunities to probe and manipulate functional properties of a wide range of materials at the nanometer scale. In particular, voltage modulated (VM) AFM techniques allow for probing electro-mechanical coupling by means of an electrically conductive nanoscale tip, enabling the unmatched exploration of local piezo- and ferro-electric behaviors amongst a long list of higher order electromechanical effects including electrostriction, flexoelectricity [2], dielectric tunability, and even ionic effects via Vegard strains etc [3-5]. This wealth of valuable information has stimulated the wide adoption of techniques such as piezoresponse force microscopy (PFM) [5] and electrochemical strain microscopy (ESM) [6] for characterization of functional nanoscale performance in materials and devices such as memory storage [7], 2D materials [8], biological systems[9], batteries [6] and fuel cells [10].

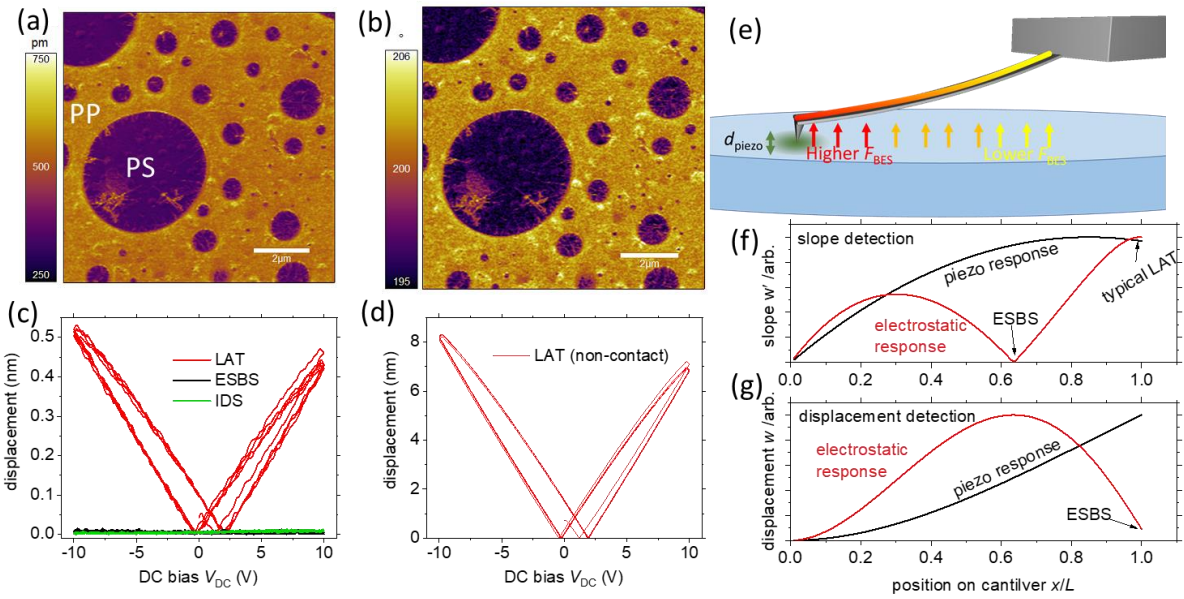

Figure 1. (a)-(d) show examples of PFM measurements on non-piezoelectric materials that exhibit false evidence of piezoelectric or ferroelectric activity. (a) and (b) show PFM imaging of amplitude and phase contrast, respectively, on a non-piezoelectric polypropylene:polystyrene polymer blend. (c) shows a PFM switching spectroscopy measurement with ferroelectric-like hysteresis on glass. Also shown in (c) is the absence of ferroelectric-like behavior when using interferometric displacement (IDS) sensing or the electrostatic blind spot (ESBS), discussed later. (d) Switching spectroscopy is repeated with the tip out-of-contact with the polymer blend, but the amplitude response still mimics ferroelectric behavior. (e) is a schematic of the mixed forcing contributions (tip forcing and long-range body electrostatic forcing) that can result in the misleading ferroelectric-like phenomena in (a)-(d). (f) and (g) are results from an Euler-Bernoulli model of the separated forcing contributions with slope or displacement detection, respectively. The position of the detection laser spot on the cantilever has a dramatic influence on the relative contribution of the desired piezo response and the undesired electrostatic response to the total cantilever amplitude. For both slope and displacement detection, ESBSs exist where the signal is sensitive to the piezo response, but insensitive to the long-range electrostatic response. More specifically, for displacement detection, the ESBS occurs near the end of the cantilever ($x/L \approx 1$), while for slope detection (used in most commercial AFMs), the ESBS occurs closer to the center ($x/L \approx 0.63$ with the chosen model parameters).

In PFM and ESM, the high-precision force and position control of the AFM allows for detection of local electromechanical deformation that arises from the material strain induced by an electrical bias applied between the tip and the sample surface. (Note penetration of the bias-induced electric field into the sample is necessary for success of the PFM measurement.) The vertical component of the material strain causes vertical displacement of the tip, resulting in an end-loaded flexural bending of the cantilever. The applied bias typically has an AC component, which can be synchronized to a lock-in amplifier that reads the cantilever bending signal from the optical beam deflection (OBD) detection system in the AFM. Because of its sensitive detection of bias-induced strain, PFM enables the non-destructive visualization and control of ferroelectric nanodomains, as well as direct measurements of the local physical characteristics of ferroelectrics, such as nucleation bias, piezoelectric coefficients, disorder potential, energy dissipation, and domain wall dynamics [5]. Despite the broad insights into ferroelectric phenomena provided by PFM, the traditional method is still plagued by artifacts that may give a misleading picture of the ferroelectric properties of a given sample [11]. As shown in Fig 1a,b, samples such as the polypropylene (PP):polystyrene (PS) polymer blend (where PP is the matrix and PS are the inclusions) can exhibit false PFM amplitude and phase contrast between "apparent" ferroelectric domains. Despite exhibiting a "calibrated" PFM amplitude nearly an order of magnitude larger than well-known ferroelectrics like lithium niobate, the sample is certainly neither piezo- or ferro-electric. It has been found empirically that most solid materials, regardless of their piezoelectric properties, will show a finite measurable response in PFM, solely due to parasitic signal contributions. Likewise, DC bias ramps on non- ferroelectric samples (as shown for glass in Figure 1c) also show apparent hysteresis that is nearly indistinguishable from true polarization switching hysteresis. This false ferroelectric hysteresis has been shown to occur even when the tip is not in contact with the surface (Figure 1d), confirming the non-electromechanical origin of the signal [12]. Ubiquitous artifacts like those demonstrated in Figure 1a-d have directly led to a rise in reports of electromechanical coupling, and sometimes false reports of ferroelectricity in materials in which ferroelectricity is absent and even forbidden (e.g., centrosymmetric monocrystalline materials)[11,13,14].

Of the various types of artifacts that affect PFM measurements, the strongest is long-range, so-called body electrostatic (BES) forces that exist between the cantilever and sample, Fig 1e. These BES forces arise from the electrostatic potential difference and the capacitive gradient between the cantilever and sample surfaces, and are present in most PFM and more generally VM-AFM experiments. The BES forces are linearly proportional to the AC bias voltage between tip and sample; thus, they scale proportionally with the desired measurand, the inverse piezo response. The magnitude of the BES force is also proportional to the total DC potential difference between the tip and sample, which is equal to the sum of the built-in "contact potential difference" and the applied DC bias. Thus the magnitude of the BES force can vary significantly in studies that require modulation of the DC bias, such as domain writing and investigations of hysteresis using switching spectroscopy [15]. In normal operation, the BES forces generate a bending response in the cantilever that the AFM cannot distinguish from the bending response to the AC bias induced normal strain in the sample (i.e., the true PFM signal). Overall, eliminating the influence of BES force in PFM is essential to improve the veracity and reliability of the measurement and to attain improved understanding of nanoscale ferroelectric phenomena.

*Reduction of Body Electrostatic Artifacts in PFM*

Upon recognizing the importance of BES artifacts to the (mis)interpretation of PFM measurements, multiple researchers have sought to mitigate the artifacts' influence. Two broad approaches to BES mitigation have been tried. In the first approach, the electrostatic force itself is reduced, such that it can no longer affect the observed measurands. An example of the first approach utilizes tall tips that place the cantilever-body further from the sample-surface, taking advantage of the distance-squared decay in electrostatic force[16]. Similarly, Hong and Shin translated the cantilever body to overhang the sample edge, leading to a reduced electrostatic force on the overhanging portion of the cantilever [17]. However, the most proximal portion of the cantilever still overlaps the sample and contributes to the artifact, the artifact magnitude varies as the scan progresses further onto the sample, and the total amount of scannable sample-area is limited. Finally, it has been demonstrated that it is possible to apply a DC bias which cancels the contact potential difference, producing a null DC potential difference and eliminating the electrostatic force[18]. In principle this approach can provide a BES-free result. However, the nulling bias can be difficult to determine, it can vary with tip location on the sample and with time, and the approach precludes techniques such as switching spectroscopy which require modulation of the DC bias. In connection with the DC bias approach, it should be emphasized that when the tip is in contact with a piezoelectric material, the measured cantilever bending is a sum of responses to the bias-induced sample strain and the BES force. Hence, adjusting the DC bias to null the measured response is equivalent to nulling the sum of the "piezoelectric" (bias-induced strain) and "electrostatic" (BES force) responses. Thus, it is not equivalent to nulling either response by itself.

In the second approach to mitigate BES artifacts, experimental parameters are chosen such that electrostatic forces are still present, but their influence on the detected signal is small compared to the inverse piezo response. Early models and measurements revealed that operation with cantilevers with relatively high spring constant can lessen the influence of the electrostatic forces compared to the electromechanical displacement [19]. Operation with very stiff cantilevers is not always desirable, and can counteract a major benefit of the AFM – force precision. Macdonald et al showed that, in contact resonance PFM (CR-PFM) experiments, higher-order (e.g. >1) contact resonance eigenmodes of the cantilever selectively become orders of magnitude less sensitive to the electrostatic force than the electromechanical force, compared to the lowest eigenmodes or quasistatic vibrations. Thus, higher mode order CR-PFM methods improve sensitivity to small electromechanical displacements, but quantification of the PFM displacement signal is challenging due to a difficult-to-measure, contact-dependent volt to nanometer optical lever sensitivity (OLS) [20,21]. Labuda et al [22] advanced artifact-free quantitative PFM by replacing the traditional slope-sensitive optical beam deflection (OBD) system of the AFM with an interferometric detection system (IDS). When the IDS beam is placed directly above tip, it senses the normal displacement of the tip, which is dominated by the underlying electromechanical strains in the sample. Despite the benefits of the interferometric method, it is limited to higher frequency operation (>10 kHz) and thus cannot fully replace the OBD system in basic AFM operation. Furthermore, IDS necessitates significant, expensive customization of the underlying AFM to introduce the required optical components; indeed, customization for IDS may be physically impossible with many of the AFMs currently in the field.

Here, we demonstrate a universal approach for performing quantitative PFM, and VM-AFM more generally, that is free of BES artifacts. This is achieved by positioning the slope sensitive OBD spot at a position along the cantilever where the bending induced by the distributed electrostatic force has no influence on the local bending slope of the cantilever[23], making the OBD response electrostatically blind. We show that this method allows for accurate imaging of nanoscale ferroelectric domains, quantitative determination of piezoelectric coupling coefficients, and unambiguous separation of true ferroelectric domain switching from non-ferroelectric hysteresis artifacts. As such, electrostatic blind spot PFM (ESBS-PFM) overcomes the challenges which have plagued PFM techniques for over 3 decades and hence enables major advances in materials characterization and exploration by PFM. Importantly, our method is universally applicable, compatible with all existing AFMs, and can be easily implemented without the need for expensive or complicated additional equipment or software. As such, we believe the adoption of ESBS detection in PFM and ESM will improve the accuracy, repeatability, comparability with theoretical models and inter-laboratory agreement of quantitative measurements of material properties such as piezoelectric coupling coefficients (nm/V).

**Results and Discussion**

*Prediction of the Electrostatic Blind Spot in PFM*

By modeling the cantilever vibration along its entire length, rather than just at the tip, existing PFM measurements can be better understood and opportunities for improved measurement are apparent. Fig 1f,g illustrates modeling the cantilever vibration in terms of slope $w'$ and displacement $w$, as would be detected by OBD and IDS, respectively. The response contributions are shown for both piezo and BES forces. Due to the height of the tip and the base-to-tip tilt of the cantilever, the electrostatic force is greatest near the tip where the cantilever is closest to the sample, and the force decays exponentially towards the cantilever base. In the quasi-static regime, at frequencies well below the cantilever's first resonance frequency, the bending response of the cantilever is a result of a linear superposition of the applied forces. Thus, the local bending induced by BES force, and the local bending induced by inverse piezoresponse of the sample, can be added (or subtracted if they are out of phase). Because of the tip-sample coupling, the BES force results in a bending deformation with maximum displacement at some location back from the tip. The piezo-response creates a direct loading on the AFM tip, and the cantilever bends accordingly. When sensing slope, positioning of the OBD laser at the tip (Laser At Tip, or LAT) results in a signal where both the BES contribution and the PFM contribution are close to their maxima, with total amplitude dictated by the weighted sum of both. Therefore, in the presence of significant BES, OBD-LAT detection cannot accurately determine the PFM surface displacement. Also, because the combined BES and PFM contributions result in a vibrational shape that is not consistent with force versus displacement or Brownian motion OLS calibration, the measured amplitudes are essentially arbitrary. In contrast to the OBD-LAT detection, as shown in the displacement plot (Fig 1g), IDS with LAT detection position results in a maximum of the piezo signal and a negligible BES signal. This explains the success of IDS in suppressing BES artifacts when the laser is precisely positioned [12]. For IDS, the LAT detection position serves as an electrostatic blind spot (ESBS).

Notably, an interferometer is not necessary to obtain the benefits of artifact-free, quantitative ESBS-PFM. Rather, the BES contribution has a null in slope, the ESBS, at the maximum of the BES induced displacement. For the PFM induced bending at the tip, the slope change is measurable at all locations along the cantilever, including the ESBS. The ESBS location is independent of the magnitude of either the BES force or sample electromechanical strain. We can therefore place the OBD laser at the ESBS and detect PFM displacements in the absence of any BES artifact. Furthermore, the volts to nanometer OLS calibration performed by standard force versus distance spectroscopy can precisely calibrate the OBD signal, enabling artifact free quantification of the sample surface displacement due to electromechanical strains.

*Experimental determination and validation of the Electrostatic Blind Spot*

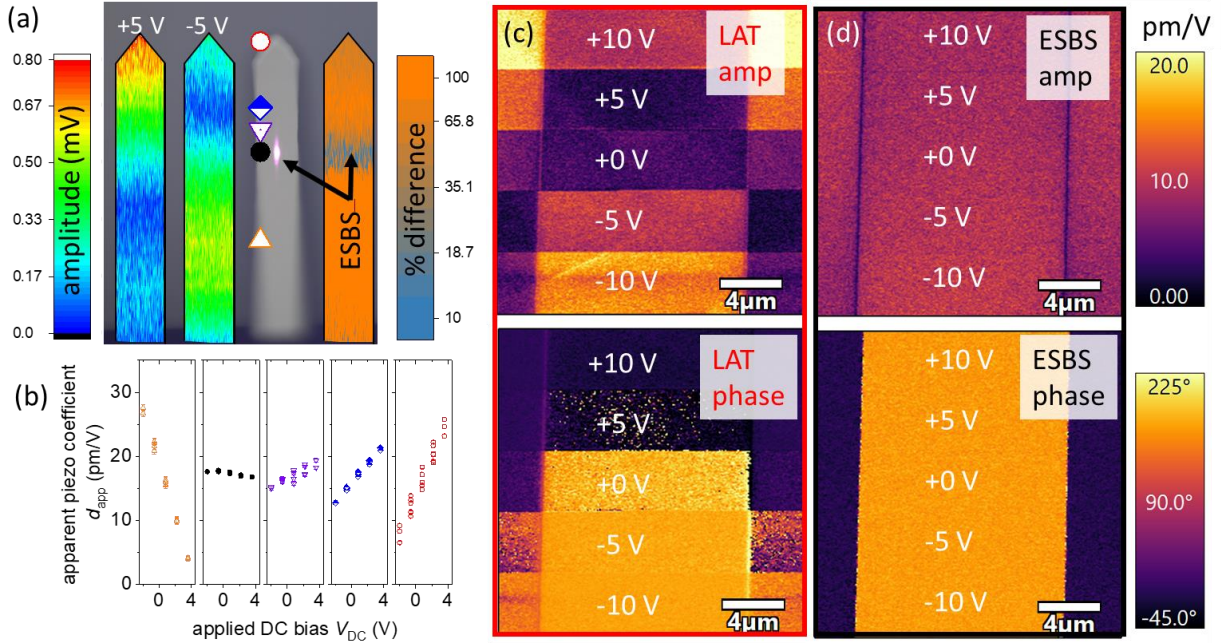

Figure 2: (a) Demonstration of the ESBS by spatial spectrogram cantilever profiling. The ESBS is determined from the minimum in the difference between positive and negative DC bias responses. (b) DC bias dependence of apparent piezo amplitude $d_{app}$ for 5 laser positions at the tip, ESBS, close to the ESBS, and towards the base. (c) PFM images on PPLN with OBD laser at the tip. DC bias is varied between +10 V and -10 V during the scan, while the amplitude and phase of the cantilever are measured. (d) same, but with laser at ESBS, showing much less effect of DC bias, which indicates reduced sensitivity to the BES force. In (c) and (d), the left and right sides of the image are referred to as domain 1 (parallel) and the center column of the images is referred to as domain 2 (anti-parallel).

Experimentally, numerous approaches exist to precisely find the ESBS. Based on the modeling results [23,24], we typically guess an ESBS position x/L≈0.6.) We can then iteratively adjust the OBD laser position and verify where the amplitude sensitivity to a variation in DC bias is minimum. We can also adjust the laser position until oppositely poled domains are equal in amplitude (assuming prior knowledge that the opposite domains should exhibit equal coupling coefficient) on a piezoelectric sample that produces strong electrostatic forces. Finally, we can engage the tip on a non-ferroelectric and non-piezoelectric sample and adjust the laser position until the minimum amplitude is observed – notably, the magnitude of the contact potential difference can vary significantly between samples, as long as the relative distribution of the BES force stays equivalent. To visually map the ESBS we employed a spatial spectrogram mapping capability in our AFM instrument[25-27]. In Fig 2a, the cantilever was brought into contact with a periodically poled lithium niobate (PPLN) substrate, then the feedback gain was disabled, fixing the extension of the z-piezo. The OBD laser was placed at 50 evenly spaced locations on the cantilever with an AC bias of 5 V. The DC voltage was then varied between –5 V and +5 V at each laser location, and the amplitude of the vibration was recorded at frequencies from 10 kHz to 30 kHz, far below the contact resonance frequency. The amplitudes are overlaid in accordance with position along the cantilever length, resulting in spectrograms of the vibrational shape of the cantilever for the positive and negative DC bias conditions. By taking the relative difference of the spectrograms at positive and negative bias we obtain a map of the amplitude dependence on electrostatic force variations (i.e. DC bias). The minimum in this difference-spectrogram indicates the ESBS, which is also represented in the optical micrograph of the cantilever.

In Figure 2b, the cantilever is brought into contact with an iron-doped lithium niobate sample. This sample is similar to PPLN, except it is not periodically poled, and its electrical conductivity is several orders of magnitude higher ( ≈ $10^{-10}$ (ohm cm) $^{-1}$ vs. ≈ $10^{-15}$ (ohm cm) $^{-1}$). As a result of the higher electrical conductivity, the LiNB:Fe sample dissipates surface charges better than PPLN, resulting in a lower tip-sample potential difference and lower BES force at small DC bias. The OBD laser was positioned at 5 spots along the cantilever and the sensitivity to varying DC bias and hence varying BES force was determined. As expected for capacitive forces, the amplitudes vary linearly with DC bias. For the non-ESBS laser positions, the slope $|dd_{app}/dV_{DC}|$ ranges from 0.6 pm/V to >4 pm/V, with the 0.6 pm/V DC bias artifact occurring less than 10 % of the cantilever length away from the ESBS. In comparison, the slope at the ESBS was <0.2 pm/V, and we expect that even an even smaller slope could have been obtained using the most precise ESBS location methods.

Figure 2c,d shows PFM amplitude and phase images on PPLN obtained with the OBD laser located at the cantilever tip and the ESBS, respectively. Between the oppositely poled domains in PPLN, we expect identical amplitude, with a 180° phase shift. A range of coupling coefficients for lithium niobate have been reported in the literature, with most reports in the range of 6 pm/V to 23 pm/V.[28] For measurements performed with the OBD laser at tip (LAT), across the ± 10 V range, the apparent coupling coefficient of domain 1 varies from 2 pm/V to 22 pm/V with an average of 9.3 pm/V ± 6.2 pm/V and the amplitude for domain 2 varies from 2 pm/V to 16 pm/V with an average of 8.3 pm/V ± 4.3 pm/V. The contrast between domains is as large as 380 %, and the mean contrast is 61 %. Likewise, the phase shift between domains over ±10 V ranges from as small as 2° to as much as 192°. In contrast, at the ESBS across the same voltage range, the coupling coefficient of domain 1 varies between 6.9 pm/V and 12.5 pm/V with an average of 9.1 pm/V ± 0.9 pm/V. For domain 2, the variation is between 7.7 pm/V and 11.3 pm/V with an average of 9.4 pm/V ± 0.8 pm/V. At all values of $V_{DC}$, the phase between domains is in the range of 163° and 185° with an average of 174°, very close to the idealized expectation, even in the presence of such significant BES forces.

*Validation of ESBS-PFM by IDS-PFM*

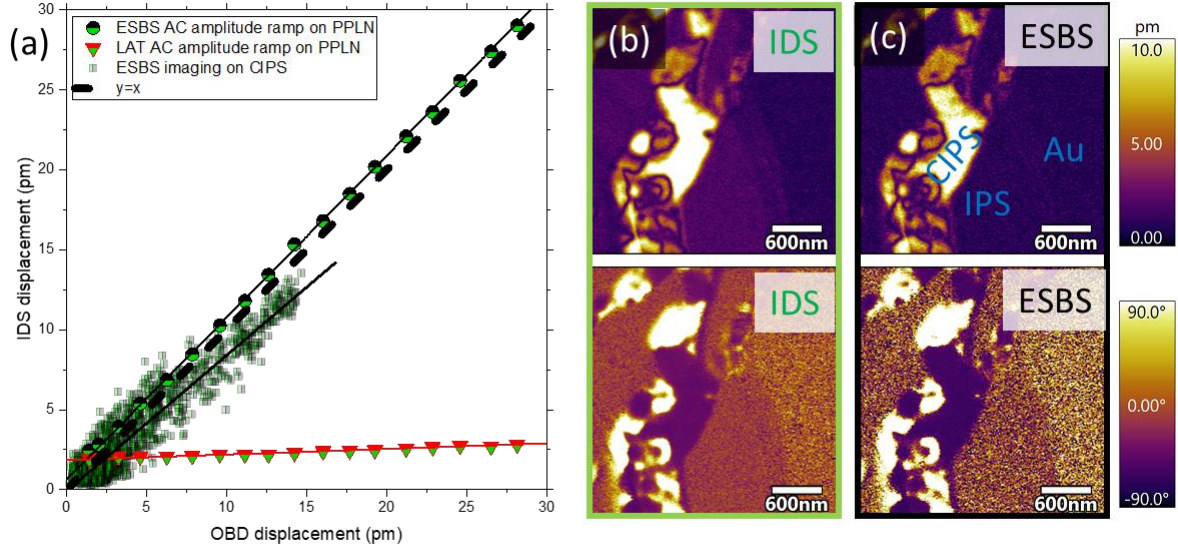

Figure 3: (a) A comparison of state of the art interferometric displacement sensing (IDS) with optical beam deflection detection with laser at the tip (LAT) and at the electrostatic blind spot, ESBS. Excellent correlation is observed between ESBS and IDS, with conversely poor correlation between LAT and IDS. Data are acquired via an AC bias ramp from 0 to 7 V on periodically poled lithium niobate and from a scan of electromechanically heterogeneous copper indium thiophosphate (CIPS). (b) and (c) show the corresponding scans on CIPS with IDS and ESBS, respectively.

To validate that OBD-ESBS PFM can accurately quantify surface electromechanical strains, OBD-ESBS results were compared to IDS-PFM results. Fig 3a shows OBD-ESBS results plotted against IDS results as AC bias was varied from 0 to 7 V while DC bias was kept at 0 on PPLN. A comparison with OBD-LAT detection is also shown. Both OBD results were separately calibrated for that OBD position based on the slope of a force versus distance curve on the same PPLN sample. The IDS and OBD-ESBS results show nearly perfect correlation, with a slope of 1.01. The only systematic deviations arise at AC bias < 0.5 V, where the OBD result approaches the noise floor yet the low-noise performance of the IDS enables continued quantification down to ≈0.25 V. Despite the laser-at-tip OBD being calibrated in the same fashion as the ESBS-OBD, it exhibits a correlation slope of only 0.03, indicating a massive underprediction of piezoresponse as the BES is out of phase with the piezoresponse. Notably, away from the ESBS, BES forces can cause overprediction or underprediction of the coupling coefficients depending on the relative phases of the different signals.

Fig 3b,c show PFM maps of amplitude and phase on ferrielectric copper indium thiophosphate $CuInP_2S_6$ (CIPS) with OBD-ESBS and IDS detection. The CIPS is a van der Waals layered material which exhibits robust ferroelectricity at room temperature having recently gained attention due to its applications in ultrathin feroic structures through exfoliation, compatibility with 2D materials for beyond-Moore electronic devices (e.g. tunnel junctions and ferroelectric field-effect transistors), as well as exhibiting significant ionic conductivity which could lead to new ferroionic states. Compared to IDS, OBD-ESBS PFM

can reliably map domains of high and low piezoresponse within the CIPS phase, as well as the precise locations of domain boundaries. These high and low regions have recently been discovered and used to rationalize a tunable quadruple-well and the co-existence of four different ferroelectric polarization states [29]. In addition, OBD-ESBS PFM correctly measures a null piezoresponse on the non-ferroelectric $InP_2S_6$ (IPS) phase. Unlike regular PFM which can contain BES signal contribution, the OBD-ESBS PFM amplitude on IPS converges to the measurement noise floor and matches that recorded on the gold electrode. The amplitude correlations on the mapped region are also represented in Fig 3a. The correlation slope is close to 1 ($m$ = 0.85), showing that quantitative agreement between IDS and OBD-ESBS can be achieved even on complex, technologically relevant samples. Lingering discrepancy between OBD-ESBS and IDS may represent small systematic errors in the IDS and ESBS state-of-the-art. For example, IDS shows a larger signal, above its noise floor, on IPS, whereas ESBS cannot detect signal above noise floor on IPS. Optimization of laser positioning for IDS and OBD-ESBS will be essential to establish which method is most accurate at its limit.

*Effects of cantilever selection*

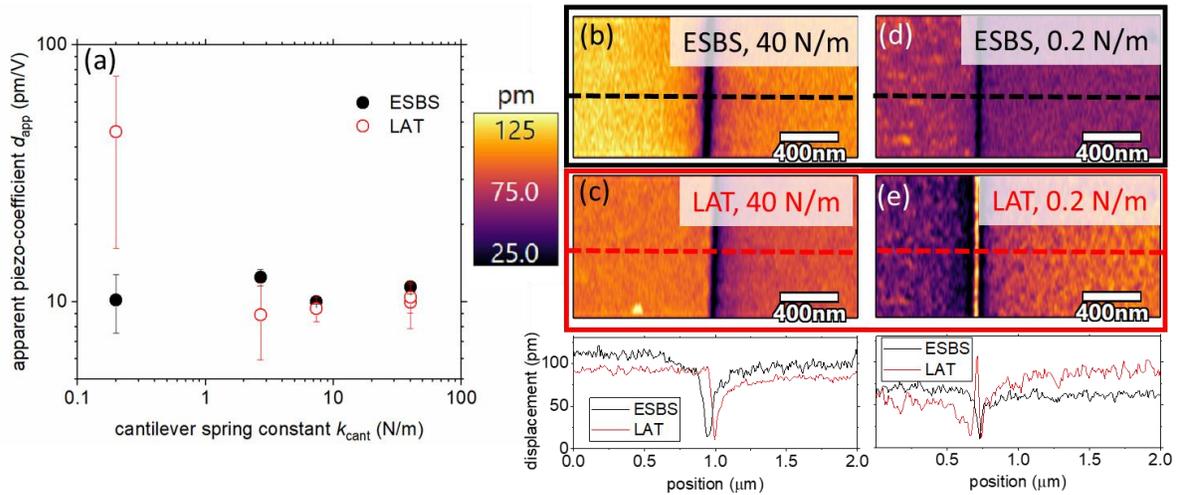

Figure 4: (a) A comparison of piezoelectric coupling coefficients measured with cantilevers of different spring constant and geometry. (b) and (d) ESBS measurements across PPLN domain boundary with 40 N/m and 0.2 N/m cantilevers, respectively. Domains show good agreement between up and down polarization, with symmetric boundary. (c) and (e) LAT measurements across same boundary showing differences in up down polarizations, asymmetric domain boundaries, artifacts at the boundary, and shifting of the boundary. Note that scan regions for a particular cantilever with different laser positions were identical to within a few nm, as confirmed by topographic features (i.e. the feature shifting is not a result of slightly different scan areas).

Choice of cantilever has long been demonstrated to have a significant effect on the veracity of PFM data. The de facto standard was to operate with a very stiff cantilever (spring constant greater than ≈40 N/m) to ensure the electrostatic artifacts were negligible. A benefit of operating with the OBD-ESBS is that it provides expanded cantilever choices. Even low spring constant cantilevers can be employed to achieve quantitative measurement. Fig 4a compares OBD-ESBS and OBD-LAT for 5 different cantilevers with nominal spring constants of 0.2 N/m, 2.8 N/m, 7.4 N/m, 40 N/m and 40 N/m. The repeated 40 N/m cases correspond with different cantilever geometries. In one case, the cantilever was 100 µm long and 50 µm wide, whereas in the other case the cantilever was 125 µm long and 30 µm wide. As shown in Fig 4a, operation with the 0.2 N/m cantilever in OBD-LAT configuration results in significant overprediction of the piezoresponse ($d_{app}$ = 46.0 pm/V ± 29.0 pm/V) as the BES is the dominant drive force. The large error bar in the OBD-LAT, 0.2 N/m result is related to significant contrast between up and down domains. For OBD-ESBS the 0.2 N/m gives a $d_{app}$ of 10.1 pm/V ± 2.6 pm/V, in good agreement with the stiff cantilevers. Although the error bar is still larger than 10 %, it is a result of measurements from spatially different areas of the sample, wherein the variation between up and down domains was less than 10 % and the variation was dominated by the new locations. For the 2.8 N/m cantilever, the agreement between laser positions improves, but the OBD-LAT still exhibits >40 % amplitude variation between domains compared to <15 % for OBD-ESBS. At 7.4 N/m, agreement between both laser positions improved significantly, with $d_{app}$ of 10.0 pm/V ± 0.2 pm/V for ESBS and $d_{app}$ of 9.4 pm/V ± 1.0 pm/V. Interestingly, the 100 µm long, 40 N/m cantilever did not exhibit an ESBS. This is attributed to the short cantilever length compared to the tip's offset from the cantilever end and the relatively large cantilever width. Combined, this geometry places too much cantilever-area forward of the tip, in proximity with the sample, for the ESBS to exist (*i.e.* the theoretical ESBS would be behind the clamp point of the cantilever for this force distribution). The change to the 125 µm long cantilever allows the ESBS to be found, and the quantified $d_{eff}$ are in good agreement with the LAT, as expected for stiffer cantilevers such as this.

While the above results suggest that absolute quantification of piezoresponse is possible with OBD-LAT and a stiff cantilever, such averaging can obscure localized artifacts that skew interpretation of underlying structure. Fig 4b,c shows maps of the up-down domain boundary in PPLN, imaged with the 40 N/m, 125 µm long cantilever. This is a configuration that would generally be thought to provide very little BES artifact in traditional PFM measurements. Indeed, the $d_{eff}$ of 10.3 pm/V from OBD-LAT in Fig 4c is close to expectations. However, investigation of the domain boundary from OBD-LAT shows an asymmetric boundary with a very "sharp" (localized) amplitude minimum. The OBD-ESBS result in Fig 4b, by comparison, shows a much more symmetric "bell-shaped" intensity distribution at the domain boundary. The amplitude minima are found to be slightly offset from one another between LAT and ESBS, indicating that electrostatic artifacts can lead to unexpected spatially correlated artifacts, even misrepresenting the location of boundaries. To qualitatively understand the domain boundary location shift and distortion in the LAT-PFM image, recall that the LAT-PFM signal is a sum of piezoelectric strain (sample surface displacement) and electrostatic components. The amplitude minima in the LAT-PFM image represent the set of locations where the two signal components cancel, which is not the same as the set of locations where the "pure" piezoelectric signal is minimum. The amplitude minima in the ESBS-PFM image do represent the set of locations where the piezoelectric strain is minimum.

For further comparison, Fig 4d,e shows the PPLN domain boundary mapped with the 0.2 N/m cantilever at ESBS and LAT laser positions. Here, LAT indicates a non-physical amplitude maximum at the boundary, whereas ESBS restores the expected near-zero amplitude at the boundary. All cases demonstrate that as the piezo-contribution decreases (e.g. at a domain boundary), the relative influence of the BES force increases, and the potential for misleading spatial artifacts also increases. Thus, accurate mapping of domain boundary geometries may be an important application of ESBS-PFM.

*Domain writing and switching spectroscopy*

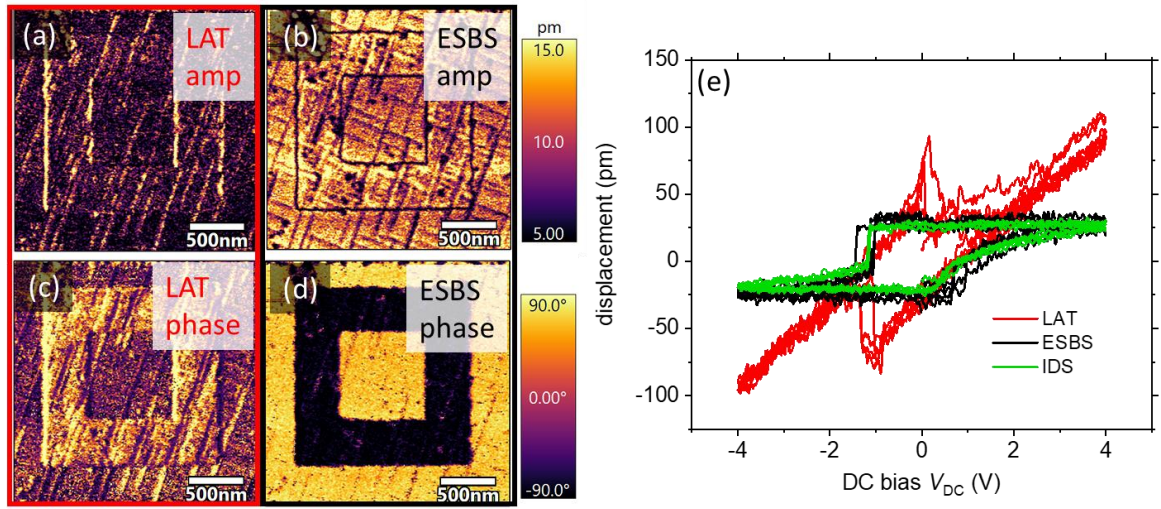

Figure 5: (a) and (c) LAT measurements of written piezoelectric domain on lead zirconate titanate (PZT) showing negligible phase contrast and underestimation of piezoelectric coupling coefficient. (b) and (d) ESBS measurements of same PZT written domain showing clear switching of polarization in phase, with higher coupling coefficient. (e) shows switching spectroscopy hysteresis loops, with an artifact-driven background slope for the LAT switching, but flat baseline for ESBS and IDS switching. Total area of the loop is also detection position and scheme dependent.

As discussed earlier, numerous PFM measurements seek to determine local electromechanical strain as a function of applied DC bias. These variations in DC bias result in variations in BES force that can mask or mimic hysteresis, as shown in Fig 1c,d. Many studies have concluded ferroelectric behavior based on these false hysteresis loops. For non-ferroelectric materials, such as batteries and fuel cells, the observed hysteresis has been assigned to ion conduction and associated volume expansions via Vegard strains. The ability to perform BES-artifact-free PFM/ESM is essential to DC bias dependence studies on complex materials.

As shown in Fig 1c, the false hysteresis on glass disappears completely when measured with IDS or ESBS. Correspondingly, Fig 5 shows results of switching experiments on lead zirconia titanate (PZT), which is expected to exhibit true ferroelectric hysteresis. Fig 5a-d show LAT and ESBS PFM scans on the PZT after domain writing at a DC voltage of ±4 V. With LAT, the coupling coefficient in the amplitude image at $V_{AC}$=1 V is underpredicted compared to when measured with the ESBS. The phase images show negligible phase contrast for LAT, but nearly 180° contrast for ESBS, indicating much better data reliability. Fig 5e shows the switching-spectroscopy measurements on the same PZT. All measurements indicate a hysteretic response, although LAT shows a strong DC dependent displacement, whereas ESBS and IDS are flat in their DC response except during switching. Interestingly, both IDS and ESBS show an asymmetry between positive bias and negative bias switching, with a sharp transition at negative bias and a gradient transition at positive bias. Because of the electrostatic background, the LAT measurement is unable to discern the true asymmetric shape of the hysteresis loop. The LAT measurement also overpredicts the area of the hysteresis loops compared to the more reliable IDS and ESBS. Loop area is widely used as a second order measure of piezoelectric responsivity, hence accurate determination is important.

**Conclusions**

We have shown theoretically and experimentally that the artifacts arising from body electrostatic forces, which have plagued voltage modulated AFM measurements for decades can be reliably eliminated by positioning the optical beam deflection laser at a location on the cantilever which is electrostatically blind, but still piezoresponse sensitive. This electrostatic blind spot can be utilized to eliminate dependence on DC bias, quantify piezoelectric coupling coefficients in a manner broadly equivalent to interferometric displacement sensing, provide interlab comparison, and expand the range of cantilever selection that is compatible with accurate, reliable VM-AFM measurements. As such, the method offers substantial benefits compared to traditional PFM operation with the laser near the tip of the cantilever, while not requiring any specialized hardware modifications to existing commercial AFMs to achieve these benefits. ESBS PFM is expected to usher in a new era of VM-AFM with more accurate portrayal of ferroelectric, piezoelectric and higher-order strain effects, with applications spanning the full range of materials that have been heretofore studied by VM-AFM methods.

**Methods**

Measurements were performed at 2 research facilities, on separate Atomic Force Microscope Instruments (Cypher, Oxford Instruments, Santa Barbara, CA). Samples were received from various vendors and collaborators, and used as is. Table 1 summarizes the experimental parameters (sample, bias voltages, cantilever selection) employed throughout the study.

Table 1: Experimental parameters

| Figure | Sample | Cantilever | AC Bias $V_{AC}$ | DC Bias $V_{DC}$ |
|---|---|---|---|---|
| 1a,b | PS-PP | Multi-75G (3 N/m) | 8 V | 0 |
| 1c,d | glass | Multi-75G (3 N/m) | 6 V | -10 V to 10 V |
| 2a | PPLN | ZEIL-PT (3 N/m) | 5 V | -5 V and 5 V |
| 2b | Fe-LiNB | ZEIL-PT (3 N/m) | 6 V | -2 V to 3.6 V |
| 2c,d | PPLN | PPP-EFM (2.8 N/m) | 6 V | -10 V to 10 V |
| 3a | PPLN | PPP-EFM (2.8 N/m) | 0 to 9 V | 0 |
| 3a-c | CIPS | PPP-EFM (2.8 N/m) | 4 V | 0 |
| 4 | PPLN | AIO Elec (0.2 N/m, 2.7 N/m, 7.4 N/m, 40 N/m), Tap 300G (40 N/m) | 10 V | 0 |
| 5a-d (writing) | PZT | CSC37 ( 0.3 N/m) | 0 | -4 V and 4 V |
| 5a-d (reading) | PZT | CSC37 ( 0.3 N/m) | 2 V | 0 |
| 5e | PZT | CSC37 ( 0.3 N/m) | 2 V | -4 V to 4 V |

To determine the ESBS, we found it most efficient to engage the sample at a desired force setpoint, then disable the force-feedback gain, fixing the position of the Z-piezo. From there, one can safely move the

OBD laser without affecting the applied force. At each new laser position, DC bias was alternated between a high and low value until the laser position with minimum sensitivity to DC bias was identified. On oppositely poled samples like PPLN additional refinements to laser position can be made while scanning across the domains with the height feedback disengaged. In this manner, the OBD laser position was adjusted until the two domains were equal in amplitude. OBD OLS was calibrated by performing a force versus distance measurement on the sample of interest. Such an approach provides high accuracy when the sample stiffness is much greater than the cantilever spring constant.

Verification of the OBD-ESBS measurements was performed with interferometric displacement sensing (IDS) via an integrated laser doppler vibrometer (Polytec GmbH, Waldbronn, Germany) on one of the AFM instruments. The IDS measurements were performed serially with the OBD measurements, on the exact same scan locations, with the same setpoint forces.

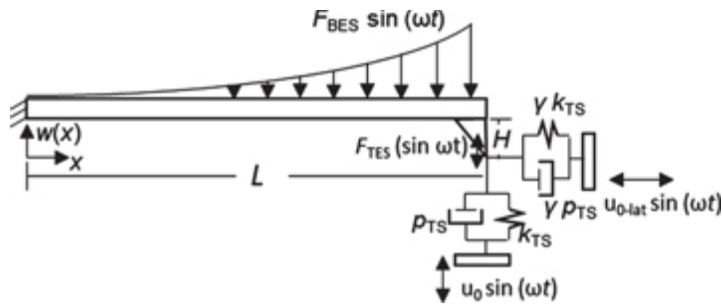

Figure 6: Euler-Bernoulli beam model used to simulate bending of cantilever.

For simulation, the cantilever was modeled as an Euler-Bernoulli beam as shown in Fig. 6. The model is described in detail in MacDonald et al[30], and is a modification of the model in [31]. The model is capable of simulating combined loading effects on the cantilever, including tip-sample electromechanical displacement $u_0$, distributed electrostatic force $F_{BES}$ and tip electrostatic force $F_{TES}$. For Fig 1f,g, the relative contributions of $u_0$ and $F_{BES}$ were varied by 6 orders of magnitude to represent the piezo-dominated and electrostatic-dominated responses. Displacement detection directly calculated local cantilever displacement $w(x)$, whereas slope detection calculates $|w'(x)|$.

**Acknowledgements**

Publication of NIST, an agency of the US government, not subject to copyright in the United States. Commercial equipment, instruments, or materials are identified only in order to adequately specify certain procedures. In no case does such identification imply recommendation or endorsement by the National Institute of Standards and Technology, nor does it imply that the products identified are necessarily the best available for the purpose.

Research was conducted at the Center for Nanophase Materials Sciences, which is sponsored at Oak Ridge National Laboratory by the Scientific User Facilities Division. This manuscript has been authored by UT-Battelle, LLC under Contract No. DE-AC05-00OR22725 with the U.S. Department of Energy. The United States Government retains and the publisher, by accepting the article for publication, acknowledges that the United States Government retains a non-exclusive, paid-up, irrevocable, world-wide license to publish or reproduce the published form of this manuscript, or allow others to do so, for United States Government purposes. The Department of Energy will provide public access to these